\documentclass[prl,aps,twocolumn,amsmath,showpacs]{revtex4}
\usepackage{bm}
\usepackage{times}
\usepackage{graphicx}
\usepackage{pifont}

\begin{document}

\title{Feedback control of nuclear hyperfine fields in double quantum dot}

\author{Wang Yao and Yu Luo}
\affiliation{Department of Physics, and Center for Theoretical and Computational Physics, The University of Hong Kong, Hong Kong, China}

\begin{abstract}
In a coupled double quantum dot system, we present a theory for the interplay between electron and nuclear spins when the two-electron singlet state is brought into resonance with one triplet state in moderate external magnetic field. We show that the quantum interference between first order and second order hyperfine processes can lead to a feedback mechanism for manipulating the nuclear hyperfine fields. In a uniform external field, positive and negative feedback controls can be realized for the gradient of the longitudinal hyperfine field as well as the average transverse hyperfine field in the double dot. The negative feedback which suppresses fluctuations in the longitudinal nuclear field gradient can enhance the decoherence time of singlet-triplet qubit to microsecond regime. We discuss the possibility of enhancing the decoherence time of each individual spin in a cluster of dots using the negative feedback control on the transverse nuclear field.
\end{abstract}
\pacs{73.21.La, 03.67.Pp, 76.70.Fz, 71.70.Jp} \keywords{}
\maketitle


\section{Introduction}
A single electron spin in semiconductor quantum dot is an attractive
candidate for a solid-state quantum bit~\cite{Loss_QDspinQC}.
Recent experiments have demonstrated the capability of reading,
writing, and controlling of single spin in different III-V quantum dot
systems, by optical
means~\cite{Imamoglu_QDSpinPrep,Steel_spinInitialization,Awschalom_spinGate,Steel_spinrabi,Yamamoto_spinRabi,Gammon_W_system},
or electric
means~\cite{1shot_r_Kouwenhoven,Jiang_FET_spin,T2star_Marcus,Kouwenhoven_singlet_triplet,Marcus_T2,Koppens_Rabi,Koppens_spinecho}.
An outstanding bottleneck towards spin-based quantum computation
has been the fast dephasing of the electron spin by the nuclear
spin environment in the III-V materials. Even at temperature as
low as $\sim100$ mK, the thermal statistical fluctuation of the
nuclear spin configurations still corresponds to a large
inhomogeneous broadening in nuclear hyperfine field, which
dephases the electron spin in a timescale of
$T_{2}^{\ast}\sim1-10$
ns~\cite{Gammon_t2star,Gurudev,Merkulov_2002,T2star_Marcus,Kouwenhoven_singlet_triplet}.
Spin echo approach can transiently remove inhomogeneous dephasing
at certain spin-echo times for an ultrashort duration equal to
$T_{2}^{*}$ time~\cite{Marcus_T2,Koppens_spinecho}, but this may
be insufficient to allow the general quantum logic controls.

In principle, the nuclear field inhomogeneous broadening can be
narrowed below its thermal value after proper procedures of
nuclear state
preparation~\cite{Burkard_measurementNarrow,Imamoglu_measurementNarrow,Loss_Narrow,
Ramon_ZamboniEffect,Levitov_selfdnp,Burkard_NuclearPreparation,Vandersypen_NuclearLocking},
and the resultant enhancement on the $T_{2}^{\ast}$ time can last
for seconds or even longer as nuclear spin relaxation is extremely
slow. This could be a solution for preparing a spin qubit with
desirable coherence properties. For optically controllable
electron spin in self-assembled dot, enhancement of $T_{2}^{\ast}$
up to microsecond timescales by nuclear state preparation has been
achieved for a spin ensemble~\cite{Greilich_nuclearfocusing}, and
for a single spin~\cite{Steel_nuclearLocking}. For electrically
controlled double quantum dot, an experimental paper has reported that a cyclic
control of the two-electron state through the
resonance between the singlet and one triplet state can enhance the
inhomogeneous dephasing time of the two-spin states to microsecond
timescale~\cite{Marcus_Zamboni}. A phenomenological model
was later proposed which shows that such control could result in negative feedback to suppress the fluctuation of
longitudinal nuclear field gradient and as a result the two-spin
dephasing time can be enhanced~\cite{Feedback}.

\begin{figure*}[t]
\includegraphics[width=14 cm, bb=84 424 538 732]{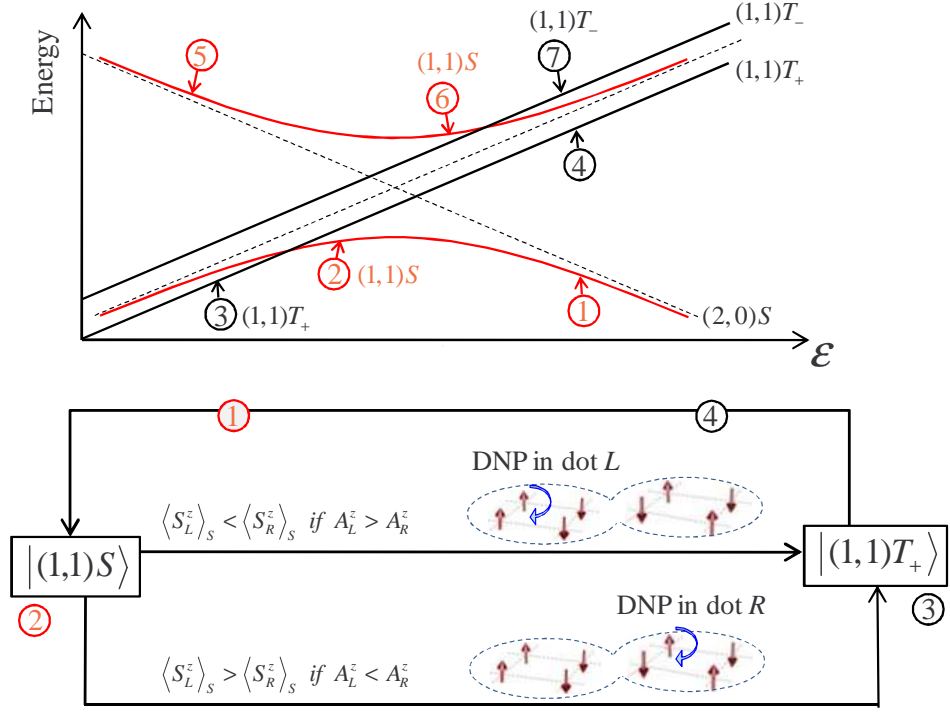}
\caption{Top: Energy level scheme of the double dot system in an external magnetic field in the $z$ direction. Bottom: feedback loop in the DNP cycle passing through the
$S-T_+$ resonance. $\varepsilon$, the detuning between the $(2,0)$ and $(1,1)$ charge
configurations, is gate-controlled. The anti-crossing of the
$(2,0)S$ and $(1,1)S$ is due to inter-dot tunneling which can be
independently controlled. We use circled numbers to denote the two-electron spin states at different values of $\varepsilon$. \textcircled{1} $\rightarrow$ \textcircled{2} $\rightarrow$ \textcircled{3} $\rightarrow$ \textcircled{4} $\rightarrow$ \textcircled{1} form a DNP control loop utilizing the $S - T_+$ degeneracy. The loop consists of: the adiabatic evolution from $(2,0)S$ to $(1,1)S$ (\textcircled{1} $\rightarrow$ \textcircled{2}); the hyperfine driven $S \rightarrow T_+$ transition which realizes the feedback on the nuclear field (\textcircled{2} $\rightarrow$ \textcircled{3}); the fast tuning of $\varepsilon$ while the system remains in the $(1,1)T_+$ state (\textcircled{3} $\rightarrow$ \textcircled{4}); re-initialization to the $(2,0)S$ by relaxation (\textcircled{4} $\rightarrow$ \textcircled{1}). The fluctuation in the nuclear field gradient
$A_{L}^{z}-A_{R}^{z}$ is suppressed (magnified) if the hyperfine
constant $\mathcal{A}$ is positive (negative). \textcircled{1} $\rightarrow$ \textcircled{5} $\rightarrow$ \textcircled{6} $\rightarrow$ \textcircled{7} $\rightarrow$ \textcircled{1} is a similar DNP control loop utilizing the $S - T_-$ degeneracy. \textcircled{1} $\rightarrow$ \textcircled{5} is realized by fast tuning of $\varepsilon$ with inter-dot tunneling switched off.} \label{Fig1}
\label{Fig1}
\end{figure*}

In this work, we present a theory for the interplay between
electron and nuclear spins near the singlet-triplet resonances of
coupled double dot in moderate external magnetic field. The transition from the singlet to a triplet in resonance can either be realized through a first order process by the electron-nuclear flip-flop term in the hyperfine coupling, or through a second order hyperfine process mediated by another detuned triplet state. We show that the quantum interference between the first order and the second order hyperfine processes leads to a feedback mechanism for manipulating the nuclear field components. The theory provides a
unified framework for realizing negative and positive feedback controls by
bringing the system through the  singlet-triplet resonances. A
non-intuitive prediction is that negative and positive feedbacks
can be realized not only for longitudinal nuclear field gradients,
but also for the nuclear field components transverse to the
external field while nuclear spins are being dynamically polarized
in the longitudinal direction. This has not been possible in other
feedback mechanisms being
explored~\cite{Levitov_selfdnp,Urbaszek_DNP,Maletinsky_DNP,Danon_DNP}.
We discuss the possibility of enhancing the dephasing time of each
individual spin in a cluster of dots using this negative feedback
control on the transverse nuclear field.

\section{Electron nuclear spin dynamics near the singlet-triplet
resonances} \label{theory}

We first briefly explain the energy level scheme which is
controlled by voltages applied to electrostatic gates in the
double dot system, as demonstrated in a number of recent
experiments~\cite{Marcus_T2,Marcus_Zamboni,Marcus_DNP,Yacoby_doubledotDNP}.
A schematic energy-level diagram with $(n,m)$ indicating the
charge occupancies of the left and right dots is shown in
Fig.~\ref{Fig1}. A total of five two-electron states are involved:
the singlet and triplet states of the $(1,1)$ charge
configuration, and the singlet state of the $(2,0)$ charge
configuration. All other states are well detuned and can be
neglected. The detuning $\varepsilon$ between the $(2,0)$ and
$(1,1)$ charge configurations is gate-controlled. The three
triplet states $(T_{-},T_{0},T_{+})$ of the $(1,1)$ configuration
are split in an external magnetic field along the $z$ direction.
In the vicinity of the $(2,0)-(1,1)$ charge degeneracy, inter-dot
tunneling causes the anti-crossing of the two singlet states
$(2,0)S$ and $(1,1)S$. Away from the $(2,0)-(1,1)$ charge
degeneracy point, the residual inter-dot tunneling results in an
exchange splitting between $(1,1)S$ and $(1,1)T_{0}$, which allows
the $(1,1)S$ state to be degenerate with $(1,1)T_{+}$ or
$(1,1)T_{-}$ at negative or positive $\varepsilon$. Dynamical
nuclear spin polarization (DNP) is highly efficient at the
$S-T_{+}$ and $S-T_{-}$ degeneracies since energy conservation can
be directly satisfied~\cite{Marcus_DNP}. Various aspects of the
DNP processes in such a system in different parameter regimes have
been studied in a recent theoretical work~\cite{DNPdoubledot}.
Here we focus on the feedback effects upon the crossing of
singlet-triplet resonances.

In the double dot system, each electron is coupled to the nuclear
spins in the same dot by the contact hyperfine interaction
$\hat{H}_{{\rm e-n}}=\sum_{i=L,R}
[\hat{S}_{i}^{z}\hat{A}_{i}^{z}+\frac{1}{2}(\hat{S}_{i}^{+}\hat{A}_{i}^{-}+\hat{S}_{i}^{-}\hat{A}_{i}^{+})
]$. Here $L$ and $R$ denote the left and the right dot respectively.
$\hat{A}_{L}^{z}\equiv\sum_{n \in L}a_{n}\hat{I}_{n}^{z}$ and
$\hat{A}_{L}^{\pm}\equiv\sum_{n \in L}a_{n}\hat{I}_{n}^{\pm}$
are the longitudinal and transverse nuclear field operators in the
left dot, and $\hat{A}_{R}^{z}$ and $\hat{A}_{R}^{\pm}$ are the
corresponding ones in the right dot. When the detuning between
$(1,1)S$ and $(1,1)T_{+}$ states are much smaller than the
electron Zeeman energy $g\mu_{B}B$ in the external field, the
off-resonance hyperfine coupling of $(1,1)S$ and $(1,1)T_{+}$ to
the other two far-detuned states $(1,1)T_{0}$ and $(1,1)T_{-}$ can
be eliminated by a standard canonical transformation
\begin{eqnarray*}
\hat{W} & = &
\exp\Big[\frac{\hat{A}_{L}^{z}-\hat{A}_{R}^{z}}{g\mu_{B}B}|T_{0}\rangle\langle
S|+\frac{\hat{A}_{L}^{+}-\hat{A}_{R}^{+}}{4\sqrt{2}g\mu_{B}B}|T_{-}\rangle\langle
S|\nonumber\\
&+&\frac{\hat{A}_{L}^{+}+\hat{A}_{R}^{+}}{2\sqrt{2}g\mu_{B}B}|T_{0}\rangle\langle
T_{+}|-\textrm{H.c.}\Big],\end{eqnarray*} and the residual second
order terms in the transformed Hamiltonian
$\hat{W}\hat{H}\hat{W}^{-1}$ become effective couplings within the
$(1,1)S-(1,1)T_{+}$ subspace. For the above perturbation expansion
to be valid, we require $g \mu_B B \gg A$ where $A \sim \sqrt{N}a$
is the characteristic magnitude of the nuclear hyperfine field.
Here $a$ is the typical coupling strength between the electron
spin and one nuclear spin, and $N$ is the number of nuclear spins
in one dot. For gate-defined III-V quantum dots, $N\sim10^{7}$ and
$a\sim\frac{\mathcal{A}}{N}\sim10^{4}$ s$^{-1}$, with
$\mathcal{A}\sim10^{11}$ s$^{-1}$ being the hyperfine constant of
the material \cite{Marcus_T2,Marcus_DNP}.

The transformed Hamiltonian in the $(1,1)S-(1,1)T_{+}$ subspace is
$\hat{H}_{{\rm S-T_{+}}}=E_{T_{+}}|T_{+}\rangle\langle
T_{+}|+E_{S}|S\rangle\langle S|+|T_{+}\rangle\langle
S|\hat{D}_{(S-T_{+})}+|S\rangle\langle
T_{+}|\hat{D^{\dagger}}_{(S-T_{+})} + |T_{+}\rangle\langle T_{+}|
\hat{F}_{T_+} + |S\rangle\langle S| \hat{F}_{S}$. $E_{T_{+}}$ and
$E_{S}$ are the energies of the singlet and triplet states
determined by the electrostatic gates and the magnetic field (see
Fig.~\ref{Fig1}). $\hat{D}_{(S-T_{+})}$, $\hat{F}_{T_+}$ and
$\hat{F}_{S}$ are operators that acts on the nuclear spin bath
\small
\begin{eqnarray}
\hat{D}_{(S-T_{+})}&\equiv&-\frac{1}{2\sqrt{2}}\left[(\hat{A}_{L}^{-}-\hat{A}_{R}^{-})+\frac{(\hat{A}_{L}^{-}+\hat{A}_{R}^{-})(\hat{A}_{L}^{z}-\hat{A}_{R}^{z})}{2g\mu_{B}B}\right],\nonumber\\
\label{d-operator} \\
\hat{F}_{T_{+}}&\equiv&-\frac{(\hat{A_L^-}+\hat{A_R^-})(\hat{A_L^+}+\hat{A_R^+})}{8g\mu_B B} , \\
\hat{F}_S&\equiv&-\frac{(\hat{A_L^z}-\hat{A_R^z})^2}{4g\mu_B B} -
\frac{(\hat{A_L^-}-\hat{A_R^-})(\hat{A_L^+}-\hat{A_R^+})}{8g\mu_B
B} .
\end{eqnarray}
\normalsize
Consider the electron spin initially on $|(1,1)S\rangle$ and the
nuclear spin bath on an arbitrary state $|J\rangle$. Near the
$(1,1)S - (1,1)T_{+}$ resonance, $\hat{H}_{{\rm S-T_{+}}}$ causes
transitions from the initial state $|(1,1)S\rangle \otimes
|J\rangle$ to the final states $|(1,1)T_+ \rangle \otimes
\hat{D}_{(S-T_{+})} |J\rangle $ and $|(1,1)S \rangle \otimes
\hat{F}_S |J\rangle$. The former corresponds to the hyperfine
driven $(1,1)S\rightarrow(1,1)T_{+}$ transition accompanied by the
simultaneous flip of a nuclear spin, and the latter corresponds to
the hyperfine mediated inter-dot and intra-dot nuclear spin
pair-flip while the electron spin state remains
unchanged~\cite{Ramon_ZamboniEffect,Yao_Decoherence}. For typical
nuclear state $|J\rangle$, the magnitude of the transition matrix
elements are: $\sqrt{\langle J| \hat{D}_{(S-T_{+})}^{\dagger}
\hat{D}_{(S-T_{+})} |J\rangle} \sim \sqrt{N} a$, and $
\sqrt{\langle J| \hat{F}_S ^{\dagger} \hat{F}_S |J\rangle } \sim
\frac{N a^2}{g \mu_B B} $. For $g \mu_B B \gg \sqrt{N}a$, we have
$\sqrt{\langle J| \hat{D}_{(S-T_{+})}^{\dagger}
\hat{D}_{(S-T_{+})} |J\rangle } \gg  \sqrt{ \langle J| \hat{F}_S
^{\dagger} \hat{F}_S |J\rangle } $, so the dominant process is the
hyperfine driven $(1,1)S\rightarrow(1,1)T_{+}$ transition.

One can similarly derive the effective Hamiltonian in the
$(1,1)S-(1,1)T_{-}$ subspace when these two states are near
resonance: $\hat{H}_{{\rm S-T_{-}}}=E_{T_{-}}|T_{-}\rangle\langle
T_{-}|+E_{S}|S\rangle\langle S|+|T_{-}\rangle\langle
S|\hat{D}_{(S-T_{-})}+|S\rangle\langle
T_{-}|\hat{D}_{(S-T_{-})}^{\dagger} + |T_{-}\rangle\langle T_{-}|
\hat{F}_{T_-} + |S\rangle\langle S| \hat{F}_{S}$, where
$\hat{D}_{(S-T_{-})} \equiv
\frac{1}{2\sqrt{2}}[(\hat{A}_{L}^{+}-\hat{A}_{R}^{+}) + \frac{(\hat{A}_{L}^{+}+\hat{A}_{R}^{+})(\hat{A}_{L}^{z}-\hat{A}_{R}^{z})}{2g\mu_{B}B}]$ determines the back-action on
the nuclear spin bath when the hyperfine driven $(1,1)S
\rightarrow (1,1)T_{-}$ transition occurs at the
$(1,1)S-(1,1)T_{-}$ degeneracy.

\section{Feedback controls of nuclear field fluctuations}

Upon the hyperfine driven $(1,1)S\rightarrow(1,1)T_{+}$
transition, the effect of $\hat{D}_{(S-T_{+})}$ is the flip down
of a nuclear spin in the left or right dot, which polarizes the
nuclear spin bath in the direction opposite to the external field.
The interesting phenomenon comes from the interference of the
leading order term with the second order term in
$\hat{D}_{(S-T_{+})}$. Consider an initial nuclear state
$|A_{L}^{z},A_{R}^{z}\rangle$ (an eigenstate of $\hat{A}_{L}^{z}$
and $\hat{A}_{R}^{z}$ with eigenvalues $A_{L}^z$ and $A_{R}^z$
respectively), $\hat{D}_{(S-T_{+})}$ brings it to the final state
\begin{eqnarray}
&\left[\left(1+\frac{A_{L}^{z}-A_{R}^{z}}{2g\mu_{B}B}\right)\hat{A}_{L}^{-}-\left(1-\frac{A_{L}^{z}-A_{R}^{z}}{2g\mu_{B}B}\right)\hat{A}_{R}^{-}\right]|A_{L}^{z},A_{R}^{z}\rangle.\label{ze}
\end{eqnarray}
Namely, there is a larger probability for the flip-down of a
nuclear spin to occur in left (right) dot if $A_{L}^{z}-A_{R}^{z}$
in the initial state is positive (negative). In either case, the
magnitude of the nuclear field gradient, is reduced. This is a
negative feedback which will reduce the magnitude of
$A_{L}^{z}-A_{R}^{z}$.

For an ensemble of identical systems or the time ensemble averaged
dynamics of a single system~\cite{Marcus_T2}, we shall consider
the ensemble average over evolutions initiated on various possible
nuclear states. $\left\langle
\left(\hat{A}_{L}^{z}-\hat{A}_{R}^{z}\right)^{2}\right\rangle $
gives the fluctuation of the nuclear field gradient in such
ensemble dynamics~\cite{average}, where $\left \langle \cdots
\right \rangle$ stands for the quantum mechanical expectation
value averaged over an ensemble of nuclear wavefunctions. We use
$\delta\left\langle
\left(\hat{A}_{L}^{z}-\hat{A}_{R}^{z}\right)^{2}\right\rangle$ to
denote the change of this fluctuation when a hyperfine driven
$S\rightarrow T_{+}$ transition has occurred. Eq.~(\ref{ze}) then
leads to
\begin{eqnarray}
\delta\left\langle
\left(\hat{A}_{L}^{z}-\hat{A}_{R}^{z}\right)^{2}\right\rangle  & =
& \frac{-2a}{|g\mu_{B}B|}\left\langle
\left(\hat{A}_{L}^{z}-\hat{A}_{R}^{z}\right)^{2}\right\rangle
+a^{2}. \nonumber \\ \label{eq:feedback-gradient}
\end{eqnarray}
It is clear that the fluctuation$\left\langle
\left(\hat{A}_{L}^{z}-\hat{A}_{R}^{z}\right)^{2}\right\rangle $
gets suppressed in the DNP cycle if
$g\mu_{B}B<\frac{2}{a}\left\langle
\left(\hat{A}_{L}^{z}-\hat{A}_{R}^{z}\right)^{2}\right\rangle $.
For a thermalized nuclear spin bath, we note that
$\frac{1}{a}\left\langle
\left(\hat{A}_{L}^{z}-\hat{A}_{R}^{z}\right)^{2}\right\rangle
\sim\mathcal{A}$, corresponding to a magnetic field of $\sim5$
T~\cite{Marcus_DNP}. The feedback effect is inversely proportional to the magnetic field. In the meantime, the perturbation treatment requires the magnetic field to be much larger than the nuclear field gradient which is of the typical value $ \sim 2 $ mT for a thermalized nuclear spin bath. Thus, we expect the feedback effects to be pronounced in a moderate magnetic field of $10 -100$ mT.

A semiclassical picture helps to understand the underlying physics
behind this feedback. The nuclear field gradient perturbs the
electron spin eigenstates in the uniform external field, which
results in an electron spin polarization gradient in the perturbed
$(1,1)S$ state. The electron spin polarization gradient in turn
determines the back-action to the nuclear spin bath upon the
electron-nuclear flip-flop, which completes the feedback loop for
manipulating the nuclear field gradient. One can further
anticipate a similar feedback effect on the transverse nuclear
field as well in the DNP process. In the presence of a transverse
nuclear field, the total effective magnetic field for the electron
is tilted from the $z$ direction, so the electron spin
polarization in the perturbed $(1,1)T_{+}$ state has a transverse
component. Upon the hyperfine driven $S\rightarrow T_{+}$
transition, nuclear spins are then polarized along the axis of the
total magnetic field, which changes the transverse nuclear field
value.

To derive the effect on the transverse nuclear field,
\textit{e.g.} along the $x$ direction, it is convenient to use the
$x$ basis for the nuclear state and we rewrite Eq.
(\ref{d-operator}), \small
\begin{eqnarray}
&&\hat{D}_{(S-T_{+})}  = \frac{1}{2\sqrt{2}}[\hat{A}_{L}^{x}-\hat{A}_{R}^{x} -\nonumber\\
&&\frac{\left(\hat{A}_{L}^{x+}+\hat{A}_{L}^{x-}+\hat{A}_{R}^{x+}+\hat{A}_{R}^{x-}\right)\left(\hat{A}_{L}^{x+}-\hat{A}_{L}^{x-}-\hat{A}_{R}^{x+}+\hat{A}_{R}^{x-}\right)}{8g\mu_{B}B}
\nonumber \\
&&+ \left(1+\frac{\hat{A}_{L}^{x}+\hat{A}_{R}^{x}}{2g\mu_{B}B}\right)\frac{\hat{A}_{L}^{x+}-\hat{A}_{R}^{x+}}{2\textrm{i}} \nonumber \\&&+\left(1-\frac{\hat{A}_{L}^{x}+\hat{A}_{R}^{x}}{2g\mu_{B}B}\right)\frac{\hat{A}_{L}^{x-}-\hat{A}_{R}^{x-}}{2\textrm{i}}]\nonumber\\
\label{eq:transverse}
\end{eqnarray}
\normalsize Consider an initial nuclear state
$|A_{L}^{x},A_{R}^{x}\rangle$, an eigenstate of $\hat{A}_{L}^{x}$
and $\hat{A}_{R}^{x}$ with eigenvalues $A_{L}^{x}$ and $A_{R}^{x}$
respectively, $\hat{D}_{(S-T_{+})}$ brings it to the final state
\small
\begin{eqnarray}
&&\hat{D}_{(S-T_{+})} |A_{L}^{x},A_{R}^{x}\rangle  =  \frac{1}{2\sqrt{2}} \bigg[ \left (A_{L}^{x}-A_{R}^{x} \right ) |A_{L}^{x},A_{R}^{x}\rangle  \nonumber \\
& +& \left(1+\frac{A_{L}^{x}+A_{R}^{x}+a}{2g\mu_{B}B}\right)\frac{\lambda |A_{L}^{x}+a,A_{R}^{x}\rangle - \lambda |A_{L}^{x},A_{R}^{x}+a\rangle }{2\textrm{i}}  \nonumber \\
&+&\left(1-\frac{A_{L}^{x}+A_{R}^{x}-a}{2g\mu_{B}B}\right)\frac{\lambda
|A_{L}^{x}-a,A_{R}^{x}\rangle - \lambda
|A_{L}^{x},A_{R}^{x}-a\rangle}{2\textrm{i}} \bigg]\nonumber\\
\label{finalstate-x}
\end{eqnarray}
\normalsize where $\lambda \equiv  \sqrt{\langle
A_{L}^{x},A_{R}^{x}| \hat{A}_{L}^{x-} \hat{A}_{L}^{x+}
|A_{L}^{x},A_{R}^{x}\rangle} \sim \sqrt{N} a$ and $|
A_{L}^{x}-A_{R}^{x} | \sim \sqrt{N} a$. Here we have neglected
terms such as $|A_{L}^{x} \pm a,A_{R}^{x} \pm a\rangle$ whose
probability in the order of $(\frac{\sqrt{N} a}{g\mu_{B}B})^2$.

A measure of $\hat{A}_{L}^{x}+\hat{A}_{R}^{x}$ in the final state
has three possible values: $A_{L}^{x}+A_{R}^{x}$,
$A_{L}^{x}+A_{R}^{x}+a$ and $A_{L}^{x}+A_{R}^{x}-a$, with the
probabilities of
\small
\begin{eqnarray*}
&&\frac{(A_{L}^{x}-A_{R}^{x})^2}{(A_{L}^{x}-A_{R}^{x})^2+
\lambda^2},\frac{\lambda^2/2}{(A_{L}^{x}-A_{R}^{x})^2+ \lambda^2}
\left(1+\frac{A_{L}^{x}+A_{R}^{x}}{g\mu_{B}B}\right),\\
&&\frac{\lambda^2/2}{(A_{L}^{x}-A_{R}^{x})^2+ \lambda^2}
\left(1-\frac{A_{L}^{x}+A_{R}^{x}}{g\mu_{B}B}\right)
\end{eqnarray*}
\normalsize respectively. If $A_{L}^{x}+A_{R}^{x}$ is initially
positive, the probability for this transverse nuclear field to
increase is larger than the probability for it to decrease.
Namely, there is a positive feedback to increase the magnitude of
the total transverse field. For the $y$ component of the nuclear
field, we have the same result. Upon a hyperfine driven
$S\rightarrow T_{+}$ transition, the change in the fluctuation of
transverse nuclear field
$\hat{\bm{A}}^{\perp}\equiv\hat{A}^{x}\bm{i}+\hat{A}^{y}\bm{j}$
for a nuclear spin ensemble is given by~\cite{average}
\small
\begin{eqnarray}
\delta\left\langle
\left(\hat{\bm{A}}_{L}^{\perp}+\hat{\bm{A}}_{R}^{\perp}\right)^{2}\right\rangle
& = & \alpha \frac{a}{|g\mu_{B}B|}\left\langle
\left(\hat{\bm{A}}_{L}^{\perp}+\hat{\bm{A}}_{R}^{\perp}\right)^{2}\right\rangle
+ \beta a^{2}, \nonumber \\ \label{eq:transverse2} \end{eqnarray}
\normalsize where $\alpha$ and $\beta$ are both positive factors
of the magnitude $\sim O(1) $. Here we note that the Larmor precession of nuclear spins in a moderate magnetic field of 10 mT is in the order of $\sim 0.1 $ MHz, which is much smaller as compared to the magnitude of the transition matrix element for the electron-nuclear flip-flop $\sim \sqrt{N} a \sim 10$ MHz. Thus, the effect of nuclear spin Larmor precession on the hyperfine driven $S\rightarrow T_{+}$ transition can be well neglected. Uniform precession of nuclear spins during other slower parts of the control cycle does not change the value of $\left\langle\left(\hat{\bm{A}}_{L}^{\perp}+\hat{\bm{A}}_{R}^{\perp}\right)^{2} \right \rangle$.

From Eq. (\ref{eq:transverse2}) and (\ref{eq:feedback-gradient}),
it is obvious that the sign of the feedback depends on the sign of
the hyperfine constant $a$. In III-V material, as the hyperfine
constants for the isotopes of Ga, In, Al and As are all
positive~\cite{Paget}, the hyperfine driven $S\rightarrow T_{+}$
transition results in a negative feedback to suppress fluctuation
in the gradient of longitudinal nuclear field and a positive
feedback to increase the fluctuation of the transverse nuclear
field.

By similar analysis, we find that the $S\rightarrow T_{-}$
transition has the opposite effects, \textit{i.e.} a positive
feedback to increase fluctuation in the gradient of longitudinal
nuclear field and a negative feedback to suppress the fluctuation
of the transverse nuclear field.

For Ge/Si double quantum
dot~\cite{SiGedoubledot_Marcus,Eriksson_SiGedoubledot}, since
$^{29}$Si and $^{73}$Ge both have negative hyperfine
constant~\cite{NMREncyclopedia}, the same control shall result in
feedbacks opposite to those in the III-V materials.

Below, we take parameters from realistic experimental systems and
evaluate the efficiency of this feedback control. In the
experiment by Reilly \textit{et al.}\cite{Marcus_Zamboni}, the
double dot is initialized on the $(2,0)S$ state when $\varepsilon$
is large and positive (see Fig.~\ref{Fig1}). By tuning
$\varepsilon$ to negative value via gate-control, the two-electron
wavefunction can be adiabatically evolved to $(1,1)S$ state. As
the evolution is slowly passed through the $S-T_{+}$ degeneracy,
the hyperfine driven $S\rightarrow T_{+}$ transition is expected
to occur with a simultaneous flip down of one nuclear spin. The
gate-control can then be rapidly brought back to the large and
positive $\varepsilon$, waiting for the double dot to be
initialized again to the $(2,0)S$ state. By repeating the above
cycle at a rate of $4$ MHz in an external field $B_{0}=10$ mT,
nuclear spins are expected to be dynamically polarized in the
direction longitudinal to the external field. The experiment is
performed at a temperature ($\sim100$ mK) many orders larger than
the nuclear Zeeman energy ($\sim0.01$ mK), so the initial nuclear
spin bath under thermal equilibrium has equal probability on every
possible spin-configuration. Before the DNP pump is applied, the
observed ensemble dephasing time $T_{2}^{*}\sim 15$ ns between the
spin states $|\uparrow\rangle_{L}|\downarrow\rangle_{R}$ and
$|\downarrow\rangle_{L}|\uparrow\rangle_{R}$ is in agreement with
the calculated fluctuation $\left\langle
\left(\hat{A}_{L}^{z}-\hat{A}_{R}^{z}\right)^{2}\right\rangle$ for
this thermal nuclear distribution~\cite{Marcus_Zamboni}.
$T_{2}^{*}$ is found to be enhanced by the cyclic
DNP pumping. After a pump time $\sim1$ s, $T_{2}^{*}$ saturates at
$\sim1~\mu$s, which indicates that $\sqrt{\left\langle
\left(\hat{A}_{L}^{z}-\hat{A}_{R}^{z}\right)^{2}\right\rangle }$
has been suppressed by a factor of $\sim70$~\cite{Marcus_Zamboni}. Measurement also
shows that this cyclic pumping establishes a steady-state nuclear
spin polarization of $\sim1\%$.

We use $\Gamma$ to denote the DNP rate, \textit{i.e.} the number
of hyperfine driven $S\rightarrow T_{+}$ transition per unit time.
Eq.(\ref{eq:feedback-gradient}) leads to the equation of motion
\small
\begin{eqnarray}
\frac{\textrm{d}}{\textrm{d}t}\left\langle
\left(\hat{A}_{L}^{z}-\hat{A}_{R}^{z}\right)^{2}\right\rangle  & =
& \frac{-2a\Gamma}{|g\mu_{B}B|}\left\langle
\left(\hat{A}_{L}^{z}-\hat{A}_{R}^{z}\right)^{2}\right\rangle
+a^{2}\Gamma, \nonumber \\ \label{eq:eom}
\end{eqnarray}
\normalsize which states that $\sqrt{\left\langle
\left(\hat{A}_{L}^{z}-\hat{A}_{R}^{z}\right)^{2}\right\rangle }$
decreases in an exponential form: $\exp(-\frac{a}{g\mu_{B}B}\Gamma
t)$. According to this exponential behavior, with a DNP rate of
$\Gamma\sim 4$ MHz, suppression of the fluctuation by a factor of
$100$ can be achieved in a timescale of $\sim 0.1$ s. To achieve
this degree of suppression, the total number of nuclear spins
flipped down by the DNP cycles shall be
$N\frac{g\mu_{B}B}{\mathcal{A}}\ln(100)\sim0.01N$, corresponding
to a nuclear polarization of $\sim1\%$. The steady-state value of
the fluctuation is reached when the two terms on the RHS of
Eq.(\ref{eq:eom}) cancel. So this residual fluctuation can be
calculated $\sqrt{\left\langle
\left(\hat{A}_{L}^{z}-\hat{A}_{R}^{z}\right)^{2}\right\rangle
}=\sqrt{a g\mu_{B}B}\approx10^{6}$ s$^{-1}$, corresponding to a
dephasing time $T_{2}^{\ast} \sim\mu$s.

By preparing the nuclear spin environment with the above cyclic
negative feedback control, the longitudinal nuclear fields in the
two dots are set to the same value regardless of the initial
state. The coherence of the physical qubit represented by the
two-spin states is therefore protected from the statistical
fluctuation of the nuclear field in each individual dot. Since
nuclear spin relaxation is extremely slow, the resultant
enhancement on the $T_{2}^{\ast}$ time can last for seconds or
even longer. It will be desirable to generalize this preparation
scheme for single spin in individual dot.

Here we argue that the negative feedback on the average transverse
nuclear field in the hyperfine driven $S\rightarrow T_{-}$
transition may potentially be utilized to enhance the dephasing
time of each single spin in a cluster of dots. As an example, we
consider a three-phase control cycle in a three-dot system. In
phase 1 of the cycle, dot 1 and dot 2 are coupled and both are
decoupled from dot 3. The two-electron state in dot 1 and 2 is
evolved in the sequence \textcircled{1} $\rightarrow$
\textcircled{5} $\rightarrow$ \textcircled{6} $\rightarrow$
\textcircled{7} $\rightarrow$ \textcircled{1} (see
Fig.~\ref{Fig1}) to realize the hyperfine driven $S\rightarrow
T_{-}$ transition. This evolution can be realized by an
experimental control similar to the one used in
Ref.~\cite{Marcus_Zamboni}.
The loop consists of four parts: (i) tuning $\varepsilon$ with inter-dot tunneling switched off (\textcircled{1} $\rightarrow$ \textcircled{5}); (ii) the adiabatic evolution from $(2,0)S$ to $(1,1)S$ (\textcircled{5} $\rightarrow$ \textcircled{6}); (iii) the hyperfine driven $S \rightarrow T_-$ transition (\textcircled{6} $\rightarrow$ \textcircled{7}); (iv) re-initialization to the $(2,0)S$ by relaxation (\textcircled{7} $\rightarrow$ \textcircled{1}).
In phase 2 (phase 3) of the cycle, we repeat the same control for
dot 2 and 3 (dot 3 and 1). The fluctuations in
$\hat{\bm{A}}_{1}^{\perp}+\hat{\bm{A}}_{2}^{\perp}$,
$\hat{\bm{A}}_{2}^{\perp}+\hat{\bm{A}}_{3}^{\perp}$ and
$\hat{\bm{A}}_{1}^{\perp}+\hat{\bm{A}}_{3}^{\perp}$ are all
reduced in this three-phase control cycle. Since these three
linearly independent vectors remain uncorrelated throughout the
control, we have
\small
\begin{eqnarray*}
\left\langle
\left(\hat{\bm{A}}_{1}^{\perp}\right)^{2}\right\rangle &=&
\frac{1}{4}\left\langle
\left(\hat{\bm{A}}_{1}^{\perp}+\hat{\bm{A}}_{2}^{\perp}\right)^{2}\right\rangle\\
&+& \frac{1}{4} \left\langle
\left(\hat{\bm{A}}_{2}^{\perp}+\hat{\bm{A}}_{3}^{\perp}\right)^{2}\right\rangle
+  \frac{1}{4} \left\langle
\left(\hat{\bm{A}}_{1}^{\perp}+\hat{\bm{A}}_{3}^{\perp}\right)^{2}\right\rangle.
\end{eqnarray*}
\normalsize Therefore, the fluctuation of the transverse nuclear
field in each individual dot gets reduced in this three-phase
control cycle. After the preparation of nuclear spin bath with the
above cycles, if the external magnetic field is rotated to one of
the transverse direction, an enhanced single spin dephasing time
$T_2^*$ in the new external field is expected.

For the qubit represented by the two-spin states
$|\uparrow\rangle_{L}|\downarrow\rangle_{R}$ and
$|\downarrow\rangle_{L}|\uparrow\rangle_{R}$, gradient in the
longitudinal nuclear field can realize a rotation about the $z$
axis. Positive feedback which increases the magnitude of the
nuclear field gradient is also of interest since a larger gradient
means a faster gate operation. A recent experiment has reported
universal quantum control of the two-spin qubit in a coupled
double-dot where the nuclear field gradient is magnified by
nuclear state preparation \cite{Yacoby_gradient}. The experiment
uses two types of controls which realize the $S\rightarrow T_{+}$
transition and the $T_{+}\rightarrow S$ transition respectively in
an external field of $\sim1$ T. Such a large external field will
deactivate the negative feedback in the $S\rightarrow T_{+}$
transition. For the $T_{+}\rightarrow S$ transition, the
back-action on the nuclear state is described by the operator
$\hat{D}_{(S-T_{+})}^{\dagger}$ which does result in a positive
feedback on the longitudinal nuclear field gradient. However, the
large external field will make it difficult to initiate the
magnification by the positive feedback. The validity of our theory
is likely to be restricted in moderate external field for both
negative and positive feedbacks, and the mechanism for the
magnification of the nuclear field gradient in this experiment is
still unclear.

\section{Summary}

For coupled double dot system in moderate external magnetic field,
we have shown that the interference of first order and second
order hyperfine processes results in a feedback mechanism for
manipulating the nuclear hyperfine fields when the two-electron
singlet state is brought into resonance with one triplet state. In
principle, the feedback controls here do not need explicit
measurement steps. Re-initialization of the control loop is simply
realized by the relaxation to the ground state $(2,0)S$ at large
positive $\varepsilon$, which can be ensured after sufficiently
long waiting time. The nondeterminacy lies in the electron-nuclear
flip-flop at the singlet-triplet resonance, but whether it has
occurred or not, the double-dot will return to the initial state
$(2,0)S$ at the end of a control cycle. The probability for the
electron-nuclear flip-flop in a single cycle corresponds to the
ratio between the actual DNP rate $\Gamma$ and the repetition rate
of the control cycles.

For negative feedback controls aimed at suppressing the nuclear
field fluctuations, spectral diffusion of nuclear hyperfine fields
caused by various
processes~\cite{Espin_HF_dipole_1_DaSSarma,Espin_HF_3_DasSarma,Yao_Decoherence,Liu_spinlong,Yao_DecoherenceControl,Cheng_holeburning}
could be a competing mechanism in determining the residual
fluctuations in the steady state. This element has not been
included in the present discussion. Further studies are needed to investigate the quantitative effects
of spectral diffusions.

WY acknowledges stimulating discussions with Xiaodong Xu, D. G. Steel, D. Gammon, and L. J. Sham. The work was supported by the Research Grant Council of Hong Kong under Grant No. HKU 706309P.


\begin{thebibliography}{41}
\bibitem{average}
Note that $\left\langle \hat{A}_{L}^{z}-\hat{A}_{R}^{z}\right\rangle$ and $\left \langle \hat{\bm{A}}_{L}^{\perp}+\hat{\bm{A}}_{R}^{\perp} \right \rangle$ are always zero under the controls being discussed.
\end{thebibliography}
\end{document}